\documentclass[twocolumn,10pt,aps,prl,preprint]{revtex4}   

\usepackage{amsmath}    
\usepackage{graphicx}   
\begin{document}
\draft
\newcommand{\ajp}{AJP}  

\title{Fermi-like behavior of weakly vibrated granular matter}
\author{Jorge E. Fiscina$^{1,2}$ and Manuel O. C\'{a}ceres$^{2,3,4}$}
\address{$^{1}$Saarland University, D-66123, Saarbr\"{u}cken Germany.\\$^{2}$Centro At\'{o}mico Bariloche, Instituto Balseiro, and $^{3}$CONICET,
\\8400 Bariloche, Argentina; and $^{4}$The Abdus Salam ICTP,34014 Trieste, Italy.}
\date{submitted to Phys.Rev.Lett., January 13$^t$$^h$, 2005}

\begin{abstract}
Vertical movement of zirconia-yttria stabilized 2 mm balls is
measured by a laser facility at the surface of a vibrated 3D
granular matter under gravity. Realizations z(t) are measured from
the top of the container by tuning the fluidized gap with a 1D
measurement window in the direction of the gravity. The statistics
obeys a Fermi-like configurational approach which is tested by the
relation between the dispersions in amplitude and velocity. We
introduce a generalized equipartition law to characterize the
ensemble of particles which cannot be described in terms of a
Brownian motion. The relation between global granular temperature
and the external excitation frequency is established.
\end{abstract}
\maketitle

A vibrated granular medium (GM) exhibits a wealth of intriguing
physical properties \cite{JNB96,HAABS97,PLMPV98,MEL03,FeMe04}.
Since energy is constantly being added to the system a
nonequilibrium steady state (s.s.) is reached \cite{Blumen94}. In
\cite{FCM04} we have studied the spectrum properties of vibrated
GM under gravity, and shown that in the weakly excited regime the
dynamics of the fluidized particles cannot be described as {\it
simple} Brownian particles, this fact leads us to the conclusion
that in order to describe the cooperative dissipative dynamics of
the GM particles, it must be done in terms of generalized Langevin
particles \cite {libro,BuC04}.

Recently Hayakawa and Hong \cite{HHo97} introduced the approach of
thermodynamics of a weakly excited granular matter, in particular vibrated
GM by mapping the nonequilibrium system with a ``Fermion like'' theory. Our
experimental conditions allow us to consider a $N$ particles system of $n$%
-rows in a cylindrical container as a 1D degenerate Fermi system. Two
experiments based on a laser were set up to investigate the occupation
dynamics at the fluidized gap of the $n$-rows GM. The first one considers a
realization $z(t)$ of one particle from the top of the container by tuning
the fluidized gap with a 1D window in the gravity direction, see Fig.1(a).
It is clear that these realizations mainly correspond to {\it macroscopic}
Fermi-like particles ({\it m}Fp) from near the ``Fermi level''. The second
one concerns the measurement after a long integration time of the mass
profile which corresponds to the Fermi-like profile, see Fig.1(b).
\begin{figure}
  \includegraphics[width=8cm]{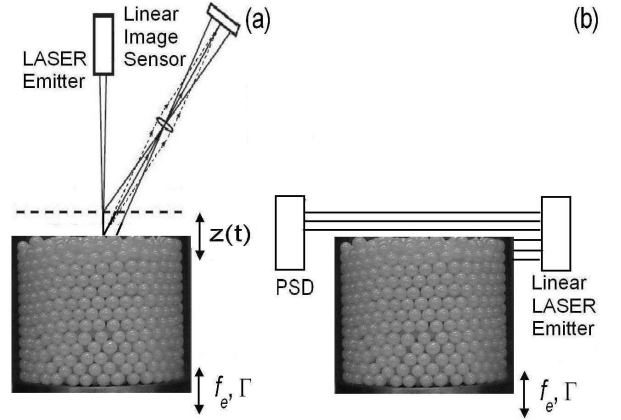}\\
  \caption{Distance measurement (amplitude realizations $z(t)$) corresponding to
$13$ layer GM ($h=$ $21$mm) with $1.99$mm $Z_{r}O_{2}-Y_{2}O_{3}$
balls in a glass container of $30$mm diameter. Set up for (a) the
single particle measurement and (b) the laser light barrier
experiment.}\label{FIG1:}
\end{figure}
By focusing on the configurational properties of an excluded
volume theory, the s.s. mass profile can be understood in terms of
a configurational maximum principle assumption. Excluded volume
interactions of the GM do not allow two grains to occupy the same
state (gravitational energy), thus the number of configurations is
$W=\prod_{i}\left[ \Omega !/N_{i}!(\Omega -N_{i})!\right] $.
Following Landau, to study a non-equilibrium system, the
maximization of $S=\ln W$ yields that the profile is $\phi
(\epsilon )=\left[ 1+Q\exp (\beta \epsilon )\right] ^{-1}\left.
\Omega \right/ mgD$ where $m,D$ are the mass and diameter of our
balls; $\beta $ is a Lagrange multiplier parameter, $Q^{-1}=\exp
(\beta \mu )$ the fugacity, and $\epsilon
=mgDs$ with $s=0,1,2,3\cdots .$ Introducing the normalization: $%
\int_{0}^{\infty }$ $\phi (\epsilon )\,d\epsilon =N,$ we get $\exp (N\beta
mgD/\Omega )=1+\exp (\beta \mu )$. Therefore the {\it zero-point} ``chemical
potential'' is $\mu _{0}=mgDN/\Omega ,$ where $N/\Omega $ is the number of
balls in an elementary {\it column} of diameter $D$. From these
considerations it is trivial to see that without vibration the centre of
mass (c.m.) is characterized by $z_{c.m.}=\mu _{0}/2mg\equiv h/2$. For
vibrated GM the Lagrange parameter $\beta $ is a non-trivial function of the
velocity fluctuations. Before going ahead, let us denote $\left. \phi
(\epsilon )\right/ N$ as the cumulative probability that the energy $%
\epsilon $ will be occupied in an {\it ideal} GM layer at the nonequilibrium
s.s. characterized by the{\it \ global} {\it temperature} $\beta ^{-1}$.

From $\phi (\epsilon )$ it is possible to calculate the c.m. expansion as a
function of $\beta $, the mean energy per particle, its square dispersion $%
\sigma _{\epsilon }^{2}$, etc. Many questions concerning the GM layer system
are still open, in fact {\it the} stochastic motion of the fluidized
particles is not entirely known \cite{GHS92}. For example it is important to
test that the spectrum of the realization $S_{z}(f)$ does not behave as
Brownian particles ($1/f^{2}$), but it has a more complex behavior, $%
1/f^{\nu }$, related to a cooperative dynamics \cite{FCM04,BuC04}. Thus an
exhaustive analysis of the realizations $z(t)$ of these {\it macroscopic }%
Fermi-like particles should be made. We have measured $z(t)$ and we
calculate the Lagrange parameter $\beta $ to show its complex relation to
the kinetic energy.

{\it Amplitude dispersion vs. the velocity dispersion}

A sinusoidal vibration is driven by a vibration plate on the GM bed (with
intensity $\Gamma =A\omega ^{2}/g$, where $A$ is the amplitude, $g$ is the
acceleration of gravity, and $\omega =2\pi f_{e}$ the frequency of the
plate). The vibration apparatus is set up by an electromagnetic shaker
(TIRAVIB$\,5212$) which allows \cite{Jorge2} for feedback through a
piezoelectric accelerometer the control of $f_{e}$ and $\Gamma $ in the
range of $10$-$7000$Hz, and $2-40g$ respectively. The control loop is
completed by an Oscillator Lab-works SC$121$ and a TIRA$\,19$/z amplifier of
$1$kw. The $n$-rows GM bed set up were $Z_{r}O_{2}-Y_{2}O_{3}$ balls with $%
D=1.99$mm and $m=26.8\pm 0.1$mgr, into a glass container of $30$ mm of
diameter with steel bottom, see Fig.1(a). The experiments were carried out
in a chamber at $1$atm of air with $5.8\pm 0.2$gr/m$^{3}$ of water vapor.
The absolute humidity was controlled by using a peltier condenser and a
control loop through a thermo-hygrometer. The humidity is of major relevance
in order to control the particle-particle and particle-wall contact forces
\cite{Jorge2,Knight95,FCM04}. Under such humidity controlled conditions, no
surface convection or convection rolls were observed in the GM, nor
rotational movement of the bed with respect to the container, which is
typical for a content of water vapor $>10$gr/m$^{3}$.

The $z(t)$ of one particle was followed in a window of $12$m\smallskip m
with a laser device by using a triangulation method, see Fig.1(a). A laser
emitter with a spot of $70\mu $m and a linear image sensor (CCD-like array)
enables a high speed measurement with $100\mu $sec sampling. The linear
image sensing method measures the peak position values for the light spots
and suppresses the perturbation of secondary peaks, which makes possible a
resolution of $1\mu $m. The shaker and the laser displacement sensor were
placed on vibration-isolated tables to isolate them from the external
vibrations, and the displacement sensor from the experiment vibrations. The $%
z(t)$ is a measure of the variations of the distance (difference) between
the particle and the sensor around the surface of the GM bed (fluidized
gap). The measured $z(t)$ without excitation reveals a white noise $<10\mu $%
m. Then our set-up effective resolution is no higher than $10\mu $m. We have
shown\cite{FCM04} that depending on the external excitation the $z(t)$ can
show from quasi non-erratic parabolas, for the movement under gravity, to
realizations of larger rugosity.

\smallskip The registers of $z(t)$ were taken with a $9354\,$C Le Croy
Oscilloscope of $500$MHz. The velocity $V(t)=\left. dz\right/ dt$ of the
{\it m}Fp was calculated numerically for $\Delta t=100\mu $sec from $z(t)$
registers. The dispersions $\sigma _{z}=\sqrt{\left\langle
z(t)^{2}\right\rangle -\left\langle z(t)\right\rangle ^{2}}$ and $\sigma
_{V}^{2}=$ $\left\langle V(t)^{2}\right\rangle -\left\langle
V(t)\right\rangle ^{2}$ were obtained from a window of $2$ second for each
pair of registers $\left\{ z(t),V(t)\right\} $. In Fig.2(a) we report $%
\sigma _{z}$ against $\sigma _{V}^{2}$ for fixed $\Gamma =10,20$ and several
$f_{e}$ from $60$-$180$Hz for GM\ beds of $h=21$mm and $12$mm.

For weakly excited GM the displacement of the fluidized particles, in the
gap, can be studied from the profile $\phi (\epsilon )$. In fact a
nonequilibrium s.s. density $P(\epsilon =mgz)$, characterizing the motion of
the fluidized {\it m}Fp, is sustained by the input of energy from the plate
colliding periodically with the GM bed; i.e., a current of particles near $%
\mu _{0}$---which is proportional to a gradient of $\phi (\epsilon )$---will
be balanced by the random input of mass coming from the periodic movement of
the plate. It is clear that $P(\epsilon )$ will be a narrow density around $%
\mu _{0},$ so we characterize the movement of the {\it m}Fp at the
Fermi-like sea by
\begin{equation}
P(\epsilon )\propto -\frac{d\phi (\epsilon )}{d\epsilon },\ \text{where }%
\epsilon =mgz=mgDs.  \label{PdeE}
\end{equation}
At the nonequilibrium s.s. the dispersion $\sigma _{z}$ can be calculated
from $P(\epsilon =mgz)$, but a rather simple and analytical expression for a
characteristic length scale $z^{*}$ can be obtained by solving $\epsilon
^{*} $ from the following consideration
\begin{equation}
qP(\mu _{0})=P(\mu _{0}+\epsilon ^{*});\ 0<q<1,  \label{qP}
\end{equation}
where $q$ is a ``cumulant'' parameter. If $P(\epsilon )$ were Gaussian the
value $q=1/\sqrt{e}$ would give the exact dispersion $\epsilon ^{*}=+\sigma
_{\epsilon }$. We have tested that our conclusions are not changed for
values $q\sim 1/\sqrt{e}$. Using $\phi (\epsilon )$ in the expression for $%
P(\epsilon )$ we get for the characteristic scale $\epsilon ^{*}$%
\begin{equation}
\exp \left( \frac{\beta \epsilon ^{*}}{2}\right) =\frac{\left( 2{\cal A}%
-1\right) +\sqrt{\left( 2{\cal A}-1\right) ^{2}-4{\cal A}q\left( {\cal A}%
-1\right) }}{2{\cal A}\sqrt{q}},  \label{eq0}
\end{equation}
where ${\cal A}=e^{\beta \mu _{0}}$, then by putting $q\sim 1/\sqrt{e}$ in (%
\ref{eq0}) it gives the amplitude dispersion $z^{*}=\epsilon ^{*}/mg$ as a
function of $\beta .$ Now the task is to determine $\beta $ as a function of
the kinetic energy of the {\it m}Fp in the fluidized gap.

If collisions were elastic, in a 1D ideal gas the equipartition theorem says
that total kinetic energy per particle is related to the Lagrange parameter
by $m\sigma _{V}^{2}/2=$ $\beta ^{-1}/2.$ Our conjecture for a weakly
excited GM is to generalize {\it the equipartition law} to
\begin{equation}
\frac{1}{2}m^{*}\sigma _{V}^{2}=\frac{1}{2}\frac{\Delta N}{N}\beta ^{-1},
\label{eq00}
\end{equation}
where $m^{*}=\delta \,m$ accounts for inelastic factors, and $\left. \Delta
N\right/ N$ is a relative factor that counts the thermodynamically
``active'' {\it m}Fp in the fluidized gap. In fact, a {\it granular gas}
since its non-Gaussian velocity distribution reveals an inelastic gas heated
in a non-uniform way, with the expected high energy tail $e^{-\text{constant}%
V^{3/2}}$.

The factor $\left. \Delta N\right/ N$ can be calculated from
\begin{equation}
\Delta N=\int_{\mu _{0}}^{\infty }\phi (\epsilon )\,d\epsilon =N\left( \frac{%
1}{\beta \mu _{0}}\ln \left[ 2e^{\beta \mu _{0}}-1\right] -1\right) .
\end{equation}
So the implicit equation to solve $\beta $ is
\begin{equation}
\frac{1}{\beta }\left( \frac{1}{\beta \mu _{0}}\ln \left[ 2e^{\beta \mu
_{0}}-1\right] -1\right) =\delta m\sigma _{V}^{2}.  \label{eq000}
\end{equation}
Note that in the high temperature limit $\beta \mu _{0}\ll 1$ and for the
elastic case $\delta =1$ we recover the equipartition theorem. This
situation is just what we have found experimentally for one steel ball in a
narrow glass cylinder \cite{FCM04}. In that experiment, when we compared the
relation $\sigma _{z}$ vs. $\sigma _{V}^{2}$, we reduced the dissipation
during the vibration and assured that there is no rotation of the ball
during its movement $z(t)$, then from energetic considerations: $mg\sigma
_{z}=\frac{1}{2}m\sigma _{V}^{2}$, predicting a line with slope $1/2g$. The
opposite situation is in the limit $\beta \mu _{0}\gg 1,$ in this case we
arrive at the {\it Low Temperature} ({\it LT}) scaling.
\begin{equation}
\beta ^{-1}\simeq \,\mid \sigma _{V}\mid \sqrt{\delta m\mu _{0}/\ln 2}.
\label{magicaBT}
\end{equation}
Due to the fact that dissipation and degrees of freedom are functions of the
external parameters, we expect that the analysis of the complex behavior of
vibrated GM will be enlightened from the study of $\sigma _{z}=\sigma
_{z}\left( \sigma _{V}^{2}\right) .$ Thus an important point would be to
test experimentally our theoretical predictions. Noting that $z^{*}=\sigma
_{z}=\epsilon ^{*}/mg$ it is simple to see that at {\it LT} (\ref{eq0}) gives
\begin{equation}
\beta \simeq 2\frac{\ln (1+\sqrt{1-q})-\ln \sqrt{q}}{mg\sigma _{z}},
\label{MágicaBT2}
\end{equation}
then using (\ref{magicaBT}) we arrive to
\begin{equation}
\sigma _{z}\simeq (\ln (1+\sqrt{1-q})-\ln \sqrt{q})\sqrt{\frac{4h\delta }{%
g\ln 2}}\mid \sigma _{V}\mid .  \label{T1}
\end{equation}
Thus we got an explicit {\it LT} formula $\sigma _{z}=\sigma _{z}\left(
\sigma _{V}^{2}\right) $ as a function of the dissipative parameter $\delta $%
, which in fact is a function of the external parameters $\Gamma $ and $%
f_{e} $.
In Fig.2(a) we report the measurement of $\sigma _{z}$
against $\sigma _{V}^{2}$ for two experimental studies of a GM bed
with $h=21$ and $h=12$ mm
at $\Gamma =10$ and $\Gamma =20$ respectively, for several $f_{e}$ from $60$-%
$180$Hz. In that figure we also show the fit with our theoretical
prediction (resolved per least squares) showing a very good
agreement for $\delta \sim 0.0009$ and $\delta \sim 0.0064$
($\Gamma =10$ and $\Gamma =20$
respectively). In Fig.2(b) we show the corresponding $\beta $ against $%
\sigma _{z}$, where the $\{\beta \}$ data set were calculated from the $%
\{\sigma _{V}\}$ experimental data set for the two experimental studies,
using (\ref{magicaBT}). By considering the mass of the $Z_{r}O_{2}-Y_{2}O_{3}
$ ball and $q\sim 1/\sqrt{e}$, we represent in Fig.2(b) the log-log plot of
the equation (\ref{MágicaBT2}), showing an excellent agreement between the
experimental data and our theory. The two experimental data sets are on the
same curve due to the fact that we use for the two experiments the same $%
Z_{r}O_{2}-Y_{2}O_{3}$ balls.
\begin{figure}
  \includegraphics[width=8cm]{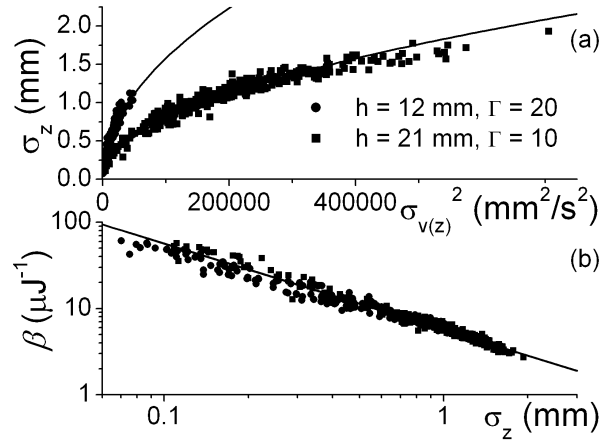}\\
  \caption{(a) $\sigma _{z}$ vs. $\sigma _{V}^{2}$, and the corresponding
fittings by using $\epsilon ^{*}$ vs. $\sigma _{V}^{2}$ solved
from (\ref {eq000}), or from (\ref{MágicaBT2}) in the {\it LT}
approximation. The excitation frequency was from $60-180$Hz. (b)
Lagrange parameter $\beta $ vs. $\sigma _{z}$.}\label{FIG2:}
\end{figure}
Note that if all stochastic realizations could be understood in
terms of a Brownian oscillating movement around $\mu _{0}$, the
$P(z)$ would correspond to $\exp \left( -z^{2}C\beta \right) $,
with $C$ a constant and $\beta ^{-1}$ proportional to the
temperature. Then we would have obtained $\sigma _{z}\propto \beta
^{-1/2}$ which is not the case reported experimentally in
Fig.2(b). Here we point out that in order to describe the spectrum
of the fluidized particles we should use a non-Markovian
description \cite{FCM04}. Unfortunately we still do not have a
time-dependent statistical description for the {\it m}Fp.
{\it Global temperature against the external excitation.}
 Eq.(\ref{magicaBT}) is the {\it LT} approximation of our generalized
equipartition theorem for a GM experiment out of equilibrium. Now
we would like to find a relation between the Lagrange parameter
$\beta $ and the external parameters characterizing the input of
energy. The maximum kinetic energy, per particle, transferred by
the oscillating plate must be
proportional to the effective mass $m^{*}$ and the dimensionless velocity $%
\left. A\omega \right/ \sqrt{gd}$, on the other hand the maximum potential
energy related to the fluidized gap is proportional to the variation of the
c.m. at the global temperature $\beta ^{-1}$. Thus we get the relation
\begin{equation}
\frac{\delta }{2}\left( \frac{A\omega }{\sqrt{gd}}\right) ^{2}=\frac{%
mg\Delta z_{c.m.}}{\mu _{0}}.  \label{Hong1}
\end{equation}
Where $\Delta z_{c.m.}=U/N-\mu _{0}/2$, with $U=\int_{0}^{\infty }$ $%
\epsilon \phi (\epsilon )\,d\epsilon .$ At {\it LT} we get $\Delta
z_{c.m.}\simeq \left. \left( \left. \pi \right/ \beta \right) ^{2}\right/
6\mu _{0},$ then
\begin{equation}
\beta ^{-1}\simeq \frac{mh\sqrt{3g\delta /d}}{\pi }A\omega ,\ h\equiv \mu
_{0}/mg  \label{Hong2}
\end{equation}
Using the relation $A\omega =\left. \Gamma g\right/ 2\pi f_{e}$ we can
transform (\ref{Hong2}) in terms of the variables that we have fixed in our
experiment. From these considerations it is trivial to see that for a given
intensity $\Gamma $ and increasing frequency $f_{e}\rightarrow \infty $ the
``temperature''$\rightarrow 0$, (i.e.$,\sigma _{z}^{2}\rightarrow 0$ and $%
z_{c.m.}\rightarrow h/2)$. In Fig.3(a) we present the behavior of the global
temperature as a function of the excitation frequency, showing an agreement
with our theoretical prediction. In Fig.3(a) a least squares fitting is also
shown giving a slope of $0.06\mu $J$^{-1}$s, while (\ref{Hong2}) gives $%
0.09\mu $J$^{-1}$s. The dispersion in Fig.3(a) is mainly introduced by the
numerical calculation of $dz/dt$.
{\it The nonequilibrium s.s. Fermi-like profile.}
\begin{figure}
  \includegraphics[width=8cm]{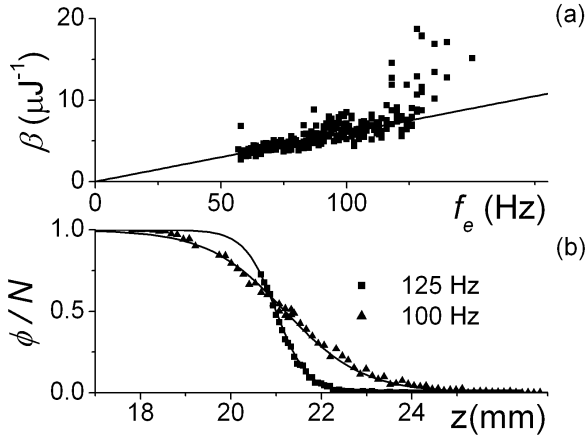}\\
  \caption{Comparison of the two experiments for a GM bed of $h=21$mm under an
acceleration $\Gamma =10$. (a) the Lagrange parameter $\beta
(f_{e})$ by measuring the realizations $z(t)$ from the top of the
container. (b) profile
$\left. \phi (z)\right/ N$ integrated from the laser light barrier during $3$%
h corresponding to $\beta =4.4\pm 0.2\mu $J$^{-1}$ for $f_{e}=100$Hz, and $%
\beta =10.5\pm 0.5\mu $J$^{-1}$ for $f_{e}=125$Hz.}\label{FIG3:}
\end{figure}
We have remarked that the profile $\left. \phi (\epsilon )\right/
N$ gives the probability that the energy $\epsilon $ will be occupied in an {\it ideal%
} GM layer at the nonequilibrium s.s. characterized by the{\it \ global
temperature} $\beta ^{-1}$. Note that $\left. \phi (\epsilon =mgz)\right/ N$
decreases monotonically with $z$ from $1$ to $0$. In fact we can write $%
\left. \phi (\epsilon )\right/ N$ as a cumulative probability
$\left. \phi (\epsilon )\right/ N=1-\int_{0}^{\epsilon }\psi
(\epsilon ^{\prime })\ d\epsilon ^{\prime }$, and interpret the
density $\psi (\epsilon )$ as associated to the fluidized gap. We
write the s.s. mass profile as
\begin{equation}
\left. \phi (z)\right/ N=\frac{(\exp (\beta mgh)-1)}{(\exp (\beta mgz)+\exp
(\beta mgh)-1)}.  \label{P1}
\end{equation}
To measure the profile (\ref{P1}) we have implemented a second experiment on
a GM bed ($h=21$mm) with a laser light barrier of $10$mm wide, Fig.1(b). The
voltage signal from the position sensitive detector runs from $0$ to $10$V,
which means a vertical window from $10$mm to $0$mm. We take measurements
every $3$s during $3$h, where the normalized frequency count integrated from
such a register is equal to the occupation number $\left. \phi \right/ N$
for $z\geq \mu _{0}/mg$ and to $1-\left. \phi \right/ N$ for $z<\mu _{0}/mg$%
. At a fixed $f_{e}$ and for $\Gamma =10$ two registers were obtained for $%
f_{e}=100$Hz and $f_{e}=125$Hz, which were integrated and normalized to get
the corresponding profile $\phi (z)/N$. In Fig.3(b) we show the profile and
our theoretical prediction (\ref{P1}). From this data we obtain the values $%
\beta =4.4\pm 0.2\mu $J$^{-1}$ for $f_{e}=100$Hz, and $\beta =10.5\pm 0.5\mu $J$%
^{-1}$ for $f_{e}=125$Hz, that we compare with the results of the
first experiment Fig.3(a). Not only the agreement is good, but
also this procedure allows a self-consistent test.

{\it Discussion.} In Fig.2(a) we show the amplitude dispersion
$\sigma _{z}$ against the velocity squared dispersion $\sigma
_{V}^{2}$ of the realizations $z(t)$. For weak amplitude the slope
$\left. \sigma _{V}^{2}\right/ \sigma _{z}$ shows a linear
behavior and the departure from a linear behavior is a clear
evidence of the complex behavior of the GM bed. This indicates
that for this regime it is necessary to introduce a description in
terms of our theory.

The corresponding global temperature $(k_{B}\beta )^{-1}$ for a fluidized
gap to occur happens to be at $T=9.2\pm 0.5\times 10^{15}$K, which means $%
f_{e}=130$Hz for $\Gamma =10$, see the {\it transition} in
Fig.3(a). Feitosa et al. measure for a dilute granular gas a range
of temperatures of the order of $T\sim 500$PK (Peta Kelvin, see
Fig.5 of Ref. \cite{FeMe04}). This range is higher than our
measurements, however it is in agreement with them since our
corresponds to a weakly fluidized GM. From \cite{FCM04} we know
that the movement of the {\it m}Fp when the fluidized gap appears
can be approximated by a Brownian motion, but this description
changes to a more complex stochastic behavior by decreasing
$f_{e}$ (for fixed $\Gamma $) when the temperature reaches $T\sim
15$PK. This global temperature should be understood, indeed, as
equivalent to an {\it order parameter} of the stochastic process
in the energy configuration. We remark that around the
fluidization transition where the stochastic dynamics start to
apply $\left. dz\right/ dt$ occurs with larger rugosity. Then a
proper description in terms of differentiable realizations $z(t)$
is well defined in the weakly excited region ($T>9$PK) where the
dynamics start to be non-Markovian. For lowertemperatures
($T\lesssim 9$PK) despite of the larger rugosity of realizations
$z(t),$ the calculation of $\sigma _{V}^{2}$ from a time-window of
$2$ second makes it reliable. This numerical calculation only
introduces, for such range, a larger dispersion of the data in the
curve of Fig.2(b), but again in good agreement with the equation
(\ref{MágicaBT2}).

 JEF thanks von Humboldt Found., ADEMAT-net,
Prof. Frank M\"{u}cklich, and Dr. Graciela C. Rodriguez. MOC
thanks the associated regime at the ICTP, and Prof. V.
Gr\"{u}nfeld for the English revision.

\end{document}